\documentclass[preprintnumbers,aps,twocolumn,floatfix]{revtex4}

\usepackage[english]{babel}
\usepackage[latin1]{inputenc}
\usepackage[dvips]{graphicx}
\usepackage{amsmath}
\usepackage{epsfig}

\begin{document}

\title{THz parametric gain in semiconductor superlattices in the
absence of electric domains}

\author{Timo Hyart}
\author{Natalia V. Alexeeva}
\author{Ahti Lepp\"{a}nen}
\author{Kirill N. Alekseev}
\email{Kirill.Alekseev@oulu.fi}

\altaffiliation{also at: Department of Physics, University of
Loughborough LE11 3TU, UK}

\affiliation{ Department of Physical Sciences, P.O. Box 3000,
University of Oulu FI-90014, Finland}

\begin{abstract}
We theoretically show that conditions for THz gain and conditions
for formation of destructive electric domains in semiconductor
superlattices are fairly different in the case of parametric
generation and amplification. Action of an unbiased high-frequency
electric field on a superlattice causes a periodic variation of
energy and effective mass of miniband electrons. This parametric
effect can result in a significant gain at some even harmonic of
the pump frequency without formation of electric domains and
corruption from pump harmonics.
\end{abstract}

\maketitle

Electrically dc-biased semiconductor superlattices (SLs)
\cite{esaki70}, which operate in conditions of single-miniband
transport regime, exhibit static negative differential
conductivity (NDC) due to existence of Bloch oscillations
\cite{esaki70}. It has been predicted that under the NDC
conditions and for a homogeneous distribution of electric field
inside nanostructure, SL should provide a strong broadband gain
with a maximum at THz frequencies \cite{ktitorov,willenberg}.
Therefore, the dc biased SL can potentially be an active element
of miniature, tunable and room-temperature operating source of THz
radiation (Bloch oscillator). However, a choice of operation point
at the part of voltage-current characteristic with negative slope
makes the SL unstable against formation of high-field domains
\cite{ktitorov,formation_domains}. The electric domains destroy
the high-frequency Bloch gain in a long SL. In order to stabilize
the electric field inside SL two new types of nanostructure design
have been recently introduced: Super-SL comprised of a stack of
short SLs \cite{savvidis04} and lateral surface SL shunted by
another SL \cite{feil}.
\par
In recent letter \cite{epl06}, we suggested to use a microwave
unbiased electric field (pump frequency $\omega$) instead of dc
bias to achieve a high-frequency gain in domainless SL devices
operating at a positive slope of voltage-current characteristic.
We theoretically proved the existence of strong negative
absorption of a sub-THz probe field with the frequency
$\omega_1=n\omega$ ($n$ is odd number). Such high-frequency gain
in SL was termed parametric. However, because SL is a strongly
nonlinear medium, odd harmonics of the pump field, $n\omega$, are
simultaneously generated \cite{ignatov76}. As a rule, the effect
of generated harmonics  blur out the weaker effect of THz
parametric gain. This problem is well-known for the parametric
amplification at microwave frequencies in Josephson junctions,
which also have strong nonlinearity \cite{param-JJ}.
\par
In this respect earlier work of Pavlovich \cite{pavlovich} merits
notice. Using approach similar to \cite{ktitorov}, he calculated
the coefficient of absorption of a weak probe field ($\omega_1$)
in SL subjected to a strong pure ac field of arbitrary frequency
$\omega$ for the case of integer or half-integer ratio of
$\omega_1/\omega$. The value of relative phase $\phi$ between the
pump and probe fields was fixed to be zero. This interesting work
has not received much attention so far \cite{comment}.
Nevertheless, simple calculations employing the original Pavlovich
formula with $\phi=0$ show that in principle gain at
$\omega_1=n\omega$ (where $n\geq 4$ is \textit{even number}) can
exist for very strong pump fields belonging to THz frequency range
$\omega\tau\simeq 1$ (the characteristic relaxation time of
electrons $\tau$ is about $100$ fs at room temperature).
Amplification at even $n$ will not be corrupted from harmonics of
the pump, because none of even harmonics can be generated in the
unbiased symmetric SL. Thus, the parametric amplification at even
harmonics is worthy of further investigation. Here the following
main questions arise: What is the physical mechanism for
amplification? What is the role of the relative phase? Is it
possible to avoid a formation of the destructive electric domains
within this scheme?
\par
In this letter, we show that a significant parametric gain at an
even harmonic arises in a superlattice due to a periodic variation
of effective masses of miniband electrons with the frequency that
is twice pump frequency or its even harmonic. We demonstrate that
the parametric gain can occur for the amplitudes and relative
phases of the pump and probe fields for which destructive electric
domains are not formed inside the superlattice. In particular, we
find that the parametric gain for small probe fields always exists
in the absence of NDC, if the pump frequency belongs to the
low-frequency part of THz range $\omega\tau\lesssim 0.45$.
\par
\begin{figure}
\includegraphics[width=0.8\columnwidth,clip=]{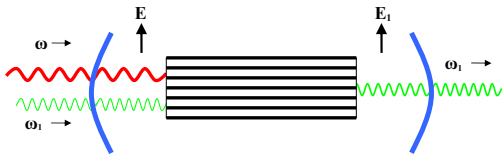}
\caption{\label{fig1}(color online) Schematic representation of SL
device. Parametric gain arises at an even harmonic
$\omega_1=n\omega$ (green online) of the unbiased pump field (red
online).}
\end{figure}

We suppose that the total electric field $E(t)$ acting on SL
electrons is the sum of the pump $E_0\cos(\omega t)$ and the probe
$E_1\cos(\omega_1 t+\phi)$ (Fig.~\ref{fig1}). Dynamics of the
electrons belonging to a single miniband is well described by the
semiclassical balance equations \cite{ignatov76,ignatov_mpl}
\begin{eqnarray}
\label{balance}
\dot{V}&=&q E(t)/m(\varepsilon)-\gamma_vV,\nonumber\\
\dot{\varepsilon}&=&q E(t)V-\gamma_\varepsilon(\varepsilon-\varepsilon_0),
\end{eqnarray}
where $V(t)$ and $\varepsilon(t)$ are the electron velocity and
energy averaged over the time-dependent distribution function
satisfying the Boltzmann equation, $q(=-e)$ is the electron
charge, $\varepsilon_0$ is the equilibrium energy of carriers,
$\gamma_v$ and $\gamma_\varepsilon$ are respectively the
relaxation constants of the average velocity and energy. The
dependence of the electron effective mass $m(\varepsilon)$ on the
energy is
$$
m(\varepsilon)=\frac{m_0}{1-2\varepsilon/\Delta},
$$
where $m_0=2\hbar^2/\Delta a^2$ is the effective mass at the
bottom of the miniband ($a$ is the SL periods and $\Delta$ is the
miniband width). Stationary solutions ($t\rightarrow\infty$) of
balance equations (\ref{balance}) have the remarkable symmetry
properties, which allow to conclude that in the unbiased case
$V(t)$ can have only odd and $\varepsilon(t)$ only even harmonics
of $\omega$ \cite{alekseev98}. Therefore, under the action of pump
field both electron energy and effective mass  vary with the
frequencies $\omega_\varepsilon=2n\omega$ ($n=1,2,3,\ldots$). For
SL placed into an external circuit (resonant cavity) with the
resonant frequency $\omega_1$ the parametric resonance condition
becomes $\omega_1=\omega_\varepsilon/2$. In terms of pump and
probe fields, one should expect parametric gain for a probe field
with the frequency $\omega_1=\omega_\varepsilon/2=n\omega$
(Fig.~\ref{fig1}).
The pump field should be strong enough in order a negative work
done on the miniband electrons in the parametric resonance can
overcome a loss due to free-carrier absorption. Parametric
interactions always depend on the value of relative phase $\phi$;
gain is expected only for some ranges of $\phi$.
\par
Here we are interested in the phase-dependent gain at even
harmonics, i.e. for $n=2,4,\ldots$. We numerically solve the
balance equations for different values of the pump $E_0$ and probe
$E_1$ amplitudes, relative phases $\phi$, for several values of
the product $\omega\tau$ ($\tau\equiv
(\gamma_v\gamma_\varepsilon)^{-1/2}$) and for several ratios
$\gamma_\varepsilon/\gamma_v$. We calculate the phase-dependent
absorption of the probe field in SL as $A=\langle
V(t)\cos(\omega_1 t+\phi)\rangle$, where time-averaging
$\langle\ldots\rangle$ is performed over the period of stationary
motion $2\pi/\omega$. Gain corresponds to a negative value of $A$.
We found that gain arises for at least one value of $\phi$ if
$E_0$ exceeds some threshold value $E_{th,n}$, which is specific
for every $n$ and also depends on $\omega\tau$.
\par
On the other hand, destructive electric domains will not be formed
inside SL if (i)  dependence of the averaged current $\langle
V(t)\rangle$ on dc bias $E_{dc}$ has a positive slope at the
working point and (ii) the frequency of the pump field is larger
than the inverse characteristic time of domain formation
$\tau_{dom}^{-1}$ ($\tau_{dom}$ is of the order of the dielectric
relaxation time) \cite{epl06}. The last condition,
$\omega\tau_{dom}>1$, can be satisfied for typical SLs with doping
$N\lesssim 10^{16}$ cm$^{-3}$ if the pump frequency
$\omega/2\pi\gtrsim 100$ GHz ($\omega\tau\gtrsim 0.1$)
\cite{epl06}. To check fulfilment of the first condition we add an
infinitesimal bias $E_{dc}$ to the pump field and numerically find
a sign of the derivative $d\langle V\rangle/dE_{dc}$ at the
working point $E_{dc}=0$. NDC and formation of domains correspond
to the negative derivative.
\par
\begin{figure}
\includegraphics[width=1\columnwidth,clip=]{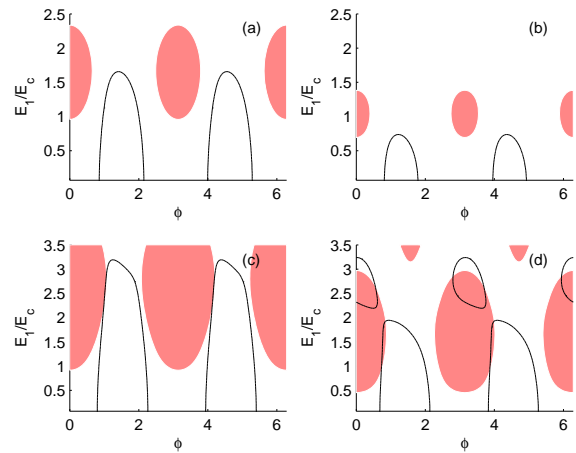}
\caption{\label{fig2}(color online) Regions of gain ($A<0$) and
regions of domain formation (red online) for the low-frequency
pump $\omega\tau=0.1$ and for (a) $n=2$, $E_0/E_c=3$; (b) $n=4$,
$E_0/E_c=3$; (c) $n=2$, $E_0/E_c=5$; (d) $n=4$, $E_0/E_c=5$.
Everywhere $\gamma_\varepsilon/\gamma_v=1$.}
\end{figure}
\begin{figure}
\includegraphics[width=1\columnwidth,clip=]{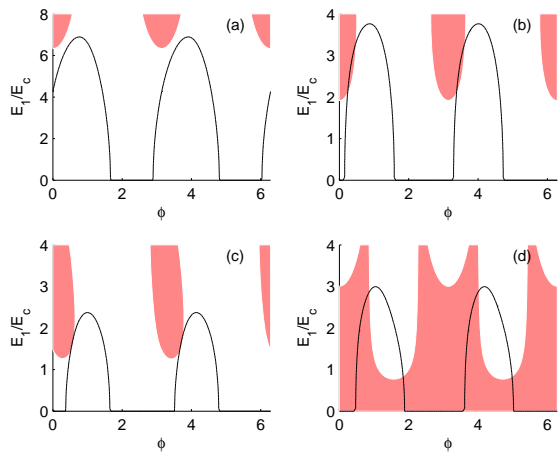}
\caption{\label{fig3}(color online) Regions of gain and domains
(red online) for the high-frequency pump fields: (a) $n=4$,
$E_0/E_c=8$, $\omega\tau=1.2$, $\gamma_\varepsilon/\gamma_v=1$;
(b) $n=2$, $E_0/E_c=4$, $\omega\tau=1.05$,
$\gamma_\varepsilon/\gamma_v=1$; (c) same as (a) but for
$\gamma_\varepsilon/\gamma_v=0.1$; (d) $n=2$, $E_0/E_c=4$,
$\omega\tau=0.75$, $\gamma_\varepsilon/\gamma_v=1$.}
\end{figure}

Figures \ref{fig2} and \ref{fig3} show the regions of gain and
domain formation in the plane $\phi E_1$ for different pump
amplitudes $E_0$ and respectively for the microwave
($\omega\tau=0.1$) and THz ($\omega\tau\simeq 1$) pump
frequencies. The field amplitudes are scaled to the Esaki-Tsu
critical field $E_c=\hbar/q a\tau$ \cite{esaki70,ignatov_mpl}. For
small $\omega\tau$ the regions of gain at second
(Fig.~\ref{fig2}a) and at fourth (Fig.~\ref{fig2}b) harmonics are
well separated from the regions of domains for all amplitudes of
probe field. In these cases the values of $E_0$ only slightly
exceed the threshold amplitudes $E_{th,2}/E_c=1.7$ and
$E_{th,4}/E_c=2.6$. With a further increase of $E_0$  these
regions of gain and domains start to overlap slightly for large
probe amplitudes (Figs.~\ref{fig2}c,d). Moreover, for relatively
high pump amplitudes or/and for gain at higher harmonics new
second-row regions of gain and domains arise for high probe
amplitudes (Fig.~\ref{fig2}d). As a rule, the second-row gain
regions strongly overlap with the regions of domains
(Fig.~\ref{fig2}d).
\par
It is worth to notice that in the quasistatic limit
$\omega\tau,\omega_1\tau\ll 1$ a series of runs demonstrated no
overlapping of the lowest NDC and gain loci in the plane $\phi
E_1$ for $n=2,4,6$ and reasonable pump strengths $E_{th,n}\leq
E_0\lesssim 5E_c$. Therefore, the domainless parametric
amplification of \textit{microwave fields} in SLs under
\textit{microwave pump} is achievable in lightly doped SLs ($N\ll
10^{16}$ cm$^{-3}$) with additionally satisfied condition
$\omega\tau_{dom}>1$ (note: $\tau_{dom}\propto m_0/N$
\cite{epl06}).
\par
Now we turn to the case of THz pump $\omega\tau\simeq 1$
(Fig.~\ref{fig3}). Complete separation of NDC and gain regions is
also possible in this case as evident from Fig.~\ref{fig3}a. This
figure also provides a good illustration for the existence of a
small-signal gain at $\phi=0$ and $n=4$ in agreement with the
original Pavlovich formula \cite{pavlovich}. We underline that it
requires a very strong pump. However, gain is still achievable for
lower pump amplitudes; in this case the range of $\phi$ supporting
gain is shifted from the point $\phi=0$ (Figs.~\ref{fig3}b,c).
Importantly, we observed no sufficient qualitative changes in the
pictures for the choice of a two different relaxation constants
$\gamma_v\neq\gamma_\varepsilon$ instead of a single constant
(e.g., cf. Figs.~\ref{fig3}b and c). Note that in the quasistatic
limit locations of NDC and gain in the plane $\phi E_1$ are
completely independent on the ratio $\gamma_\varepsilon/\gamma_v$.
\par
In all figures presented so far no NDC exists for all $\phi$ in
the limit of weak probe $E_1\rightarrow 0$. Situation is changed
drastically if the pump field itself can cause NDC for all $\phi$
(Fig.~\ref{fig3}d). It happens if $\omega\tau\gtrsim 0.45$ and
$\alpha\equiv\frac{q a
E_0}{\hbar\omega}=\frac{E_0}{E_c\omega\tau}$ is close to one of
the Bessel roots: $J_0(\alpha)=0$ (Fig.~\ref{fig3}d corresponds to
$\alpha=5.33$ that is near the second root $5.52$). Obviously, the
situation when the pump parameters correspond to the Bessel roots
should be avoided in order to reach domainless gain.
\par
In the limit of small probe field, the phase-dependent gain always
has a maximum at some optimal phase $\phi_{opt}$. We found that
the values of $\phi_{opt}$ are located \textit{at the centers} of
gain $\phi$-intervals for all $\omega\tau$ . In the quasistatic
limit $\phi_{opt}=\pi/2$ and $3\pi/2$ \cite{jap06}. For parameters
used to plot Figs. \ref{fig2}a, \ref{fig2}b, \ref{fig2}d,
\ref{fig3}a, \ref{fig3}b, and for $a=6$ nm, $\Delta=60$ meV,
$\tau=200$ fs, $N={10}^{16}$ cm$^{-3}$, the values of small-signal
gain calculated at $\phi=\phi_{opt}$ and at room temperature are
respectively $35.3$, $10.6$, $23.3$, $4$, and $34.9$ cm$^{-1}$.
The magnitude of parametric gain in SLs is no less than the
estimated Bloch gain \cite{willenberg}.
\par
Oscillator based on the parametric gain in SL should include an
external resonant cavity (circuit) tuned to $n\omega$. Our
preliminary simulations of the model, which include the balance
equations (\ref{balance}) together with the resonant cavity
equation, demonstrate that in conditions of parametric resonance
SL selects $\phi=\phi_{opt}$ among other phases of initial field
fluctuations in the cavity. Although for $\omega\tau\simeq 1$ the
computed phase of a large field in the cavity was found to deviate
already sufficiently  from $\phi_{opt}$ , it is still, as a rule,
distinctly different from the phase values supporting formation of
domains. Therefore, \textit{for the case of oscillator} partial
overlapping of the lowest gain and domain regions, like those
shown in Figs. \ref{fig2}c, \ref{fig2}d, \ref{fig3}b, will pose
\textit{no obstacle to the domainless operation} of device.
\par
SL also can be used as an active element of regenerative
parametric amplifier. In this case, the phase of pump and the
phase of an external amplified signal determine the value of phase
difference $\phi$. For operation of SL parametric amplifier it is
important to choose the pump amplitude $E_0$ and the range of
signal amplitudes $E_1$ in such a way that the corresponding
region of gain would be well separated from domains in the plane
$\phi E_1$.
\par
In summary, solving balance equations we have shown that use of ac
pump field instead of dc bias in room-temperature superlattice
devices allow to get a significant parametric gain for THz
frequencies in the absence of electric domains. In conclusion, we
would like to notice that following \cite{patane} balance
equations of the same functional form as Eq. (\ref{balance}) can
well describe a THz response of hot electrons in dilute nitride
Ga(AsN) alloys. Therefore, we suggest to explore the dilute
nitride alloys as an active media for THz parametric devices.
\par
This work was supported by Academy of Finland (grant 109758), Emil Aaltonen Foundation
and AQDJJ Programme of ESF.

\end{document}